\documentclass [11pt]{article}
\usepackage{graphicx}
\usepackage{subfigure}
\def\journal#1#2#3#4{ {#1} {\bf #2}, {#3}\  ({#4})}

\def\AmJPhys{\journal{Am.\ J.\ Phys.}}

\def\ibid{\journal{\em ibid.}}

\def\MPLA{\journal{Mod.\ Phys \ Lett. \ {\bf A}}}

\def\NPProc{\journal{Nucl.\ Phys.\ Proc.\ Suppl}}

\def\PLB{\journal{Phys.\ Lett.\ {\bf B}}}
\def\PhysRev{\journal{Phys.\ Rev.}}

\def\PRD{\journal{Phys.\ Rev.\ {\bf D}}}

\begin{document}

\newcommand{\gn}{\mbox{$\gamma_{\stackrel{}{5}}$}}
\newcommand{\adag}{a^{\dagger}}
\newcommand{\adagps}{a^{\dagger}_{p,s}}
\newcommand{\atildedag}{\tilde{a}^{\dagger}_{-p,s}}
\newcommand{\bdag}{b^{\dagger}_{-p,s}}
\newcommand{\btildedag}{\tilde{b}^{\dagger}_{-p,s}}
\newcommand{\aprime}{a^{\prime}}
\newcommand{\aprimedag}{a^{\prime \dagger}}
\newcommand{\apsbeta}{a^{\beta}_{p,s}}
\newcommand{\apsbetadag}{a_{-p,s}^{\beta\dagger}}
\newcommand{\adagbdag}{a^{\dagger}_{p,s} b^{\dagger}_{-p,s}}
\newcommand{\aps}{a^{}_{p,s}}
\newcommand{\bps}{b^{}_{-p,s}}
\newcommand{\bpsbeta}{b^{\beta}_{p,s}}
\newcommand{\bpsbetadag}{b_{-p,s}^{\beta\dagger}}
\newcommand{\Adag}{A^{\dagger}_{p,s}}
\newcommand{\Bdag}{B^{\dagger}_{-p,s}}
\newcommand{\Aps}{A^{}_{p,s}}
\newcommand{\Apsbeta}{A^{\beta}_{p,s}}
\newcommand{\Apsbetadag}{A^{\beta\dagger}_{-p,s}}
\newcommand{\Bps}{B^{}_{p,s}}
\newcommand{\Bpsbeta}{B^{\beta}_{p,s}}
\newcommand{\Bpsbetadag}{B^{\beta\dagger}_{-p,s}}
\newcommand{\ApL}{A^{}_{p,L}}
\newcommand{\AdagpL}{A^{\dagger}_{-p,L}}
\newcommand{\AdagL}{A^{\dagger}_{p,L}}
\newcommand{\AdagpR}{A^{\dagger}_{-p,R}}
\newcommand{\AdagprimepL}{A^{\prime\dagger}_{-p,L}}
\newcommand{\BpL}{B^{}_{-p,L}}
\newcommand{\ApR}{A^{}_{p,R}}
\newcommand{\BpR}{B^{}_{-p,R}}
\newcommand{\apL}{a^{}_{p,L}}
\newcommand{\akL}{a^{}_{k,L}}
\newcommand{\bpL}{b^{}_{-p,L}}
\newcommand{\apR}{a^{}_{p,R}}
\newcommand{\bpR}{b^{}_{-p,R}}
\newcommand{\BdagpL}{B^{\dagger}_{-p,L}}
\newcommand{\BdagpR}{B^{\dagger}_{-p,R}}
\newcommand{\adagpL}{a^{\dagger}_{-p,L}}
\newcommand{\adagR}{a^{\dagger}_{p,R}}
\newcommand{\bdagpL}{b^{\dagger}_{-p,L}}
\newcommand{\bdagkR}{b^{\dagger}_{-k,R}}
\newcommand{\bdagpR}{b^{\dagger}_{-p,R}}
\newcommand{\chiL}{\chi_{_L}}
\newcommand{\chiR}{\chi_{_R}}
\newcommand{\eps}{\epsilon}
\newcommand{\gnplus}{\gamma \cdot n_{_{+}}}
\newcommand{\gnminus}{\gamma \cdot n_{_{-}}}
\newcommand{\gnplusdef}{\left( \vec{\gamma} \cdot \hat{n}-\gamma_o \right)}
\newcommand{\gnminusdef}{\left( \vec{\gamma} \cdot \hat{n}+\gamma_o \right)}
\newcommand{\abab}{a^{\dagger}_{p,L}\,b^{\dagger}_{-p,L}\,a^{\dagger}_{p,R}
                   \,b^{\dagger}_{-p,R}}
\newcommand{\alphai}{\alpha_{i}}
\newcommand{\limit}{\lim_{\Lambda^2 \rightarrow \infty}}
\newcommand{\p}{\vec{p}, p_o}
\newcommand{\poprime}{p_o^{\prime}}
\newcommand{\prodps}{\prod_{p,s}}
\newcommand{\prodp}{\prod_{p}}
\newcommand{\psibar}{\bar{\psi}}
\newcommand{\psibarpsi}{ < \bar{\psi} \, \psi
            > }
\newcommand{\PsibarPsi}{ < \bar{\Psi} \, \Psi
            > }
\newcommand{\psibeta}{\psi^{}_{\beta}}
\newcommand{\psibarbeta}{\bar{\psi}_{\beta}}
\newcommand{\psidag}{\psi^{\dagger}}
\newcommand{\psidagbeta}{\psi^{\dagger}_{\beta}}
\newcommand{\psiL}{\psi_{_{L}}}
\newcommand{\psiR}{\psi_{_{R}}}
\newcommand{\Q}{Q_{_{5}}}
\newcommand{\Qa}{Q_{_{5}}^{a}}
\newcommand{\Qbeta}{Q_{5}^{\beta}}
\newcommand{\qqbar}{q\bar{q}}
\newcommand{\sumps}{\sum_{p,s}}
\newcommand{\thetap}{\theta_{p}}
\newcommand{\costhetap}{\cos{\thetap}}
\newcommand{\sinthetap}{\sin{\thetap}}
\newcommand{\thetaset}{\{ \thetap  \}}
\newcommand{\thetapi}{\thetap{}_{i}}
\newcommand{\Tomega}{\frac{\Tprime}{\omega}}
\newcommand{\pomega}{\frac{p}{\omega}}
\newcommand{\Tprime}{T'}
\newcommand{\Tprimesq}{T^{'2}}
\newcommand{\vac}{| vac \rangle}
\newcommand{\vacbeta}{| vac \rangle_{_{\beta}}}
\newcommand{\x}{\vec{x},t}
\newcommand{\xPrime}{\vec{x} - \hat{n} (t - t' ), t'}
\newcommand{\xPrimet}{\vec{x} + \hat{n} (t - t'), t'}
\newcommand{\y}{\vec{y}, y_o}

\vspace*{-.15in}

\hspace*{\fill}\fbox{CCNY-HEP-04-12}

\begin{center}
{\Large {\bf Tritium $\beta$-Decay Endpoint for a Tachyonic Neutrino  \\
		 that travels Faster than Light \fnsymbol{footnote}\footnote
	{\samepage \sl
\noindent \parbox[t]{138mm}{ \noindent
                       This work has been supported in part by a 
                       grant from PSC-BHE of CUNY. 
                           } 
}   
} 
}\\
\baselineskip 5mm
\ \\
Ngee-Pong Chang (npccc@sci.ccny.cuny.edu)\\
Department of Physics\\
The City College \& The Graduate Center of \\
The City University of New York\\
New York, N.Y. 10031\\
\  \\
October 7, 2004  \\
\end{center}

\vspace*{-.15in}

\noindent\hspace*{\fill}\parbox[t]{5.5in}{
        \hspace*{\fill}{\bf Abstract}\hspace*{\fill} \\
        {\em

	In this paper, we study the tritium $\beta-$decay in the field theory of 
	the  superluminal (faster-than-light) neutrino.  We show how the unstable transient modes
	with $\vec{q} \cdot \vec{q} < m_{\nu}^2$ lead to an excess of events at the endpoint. 
        } 
                              }\hspace*{\fill} \\             

\begin{flushleft}
\end{flushleft} 



\section{Introduction}

	Speculations that neutrinos are tachyonic have been made over the years, largely motivated by the
	negative mass-squared results from the precision measurements of the tritium $\beta$-decay end-point 
	spectrum measurements by the Mainz group\cite{Mainz} and the Troitsk group\cite{Troitsk}.  
\begin{eqnarray}
	m^2_{\nu_e}  = ( - 1.6 \pm 2.5 \pm 2.1 ) \;eV^2  && {\rm (Mainz)} \\
	m^2_{\nu_e}  = ( - 2.3 \pm 2.5 \pm 2.0 ) \;eV^2  && {\rm (Troitsk)} 
\end{eqnarray}
	While we await the next generation of even higher precision experiment from the Karlsruhe Tritium Neutrino 
	Experiment\cite{KATRIN} (KATRIN group), it is a valid question to ask if a tachyonic neutrino is even
	a self-consistent theory, and to look for possible physical consequences of such a possibility.
  
	A particularly interesting proposal was made by Chodos, Hauser, and Kostelecky in 
	1985(ref\cite{tachyon-chodos}).  They wrote down the $4$-component Dirac-like equation
\begin{equation}
	\gamma \cdot \partial \;\; \Psi (\vec{x}, t)  \;=\; -\; m \; \gamma_5 \; 
			\Psi (\vec{x}, t)   \label{chodos-equation}
\end{equation}
	where the $\gamma_5$ is crucial to making the mass tachyonic.  However, no attempt was made to quantize 
	the theory or investigate the field theoretic properties of the tachyonic neutrino.  Nevertheless, 
	they pointed out that the tritium end-point spectrum shape is sensitive to the tachyonic vs normal nature  
	of the neutrino.  

	To set the notation for our later discussion, we shall use $p, E$ for electron $4$-momenta, and reserve 
	$q, q_o$ for neutrino $4$-momenta, and refer to $x$ as the electron energy difference from its maximum
	energy
\begin{equation}
	x  \equiv  E_{max} - E
\end{equation}
	Depending on the nature of the neutrino, $E_{max}$ takes on the respective values
\begin{eqnarray}
	E_{max} \;{\rm (normal)} &=& \Delta - p^2/2M - m_{\nu} \\
	E_{max} \;{\rm(tachyon)} &=& \Delta - p^2/2M
\end{eqnarray} 
	where $\Delta$ is the mass difference between the nuclei. 

	Near the endpoint, for $x \rightarrow 0$, Chodos et al\cite{tachyon-chodos} gave the comparison
\begin{eqnarray}
	\frac{dN}{dx}(normal) &\propto & m_{\nu}  \cdot \sqrt{2m_{\nu}x }  \\
	\frac{dN}{dx}(tachyon) &\propto&   m_{\nu} \cdot   x
\end{eqnarray}
	so that the shape of the endpoint spectrum gives a clear distinction between a normal neutrino vs a 
	tachyonic neutrino.  
	
	In giving this estimate for the tachyonic neutrino, however, they have included only  physical spacelike momenta, 
	$\vec{q} \cdot \vec{q} - q_o^2 = m_{\nu}^2 $, in the decay lifetime,
\begin{eqnarray}
	\tau &=& \;\;\; \frac{1}{2} \cdot \frac{1}{(2P'_o)}\int \frac{d^3P}{(2\pi)^3(2P_o)}
	 							\frac{d^3p}{(2\pi)^3(2p_o)} \frac{d^3q}{(2\pi)^3(2q_o)} 
	 							(2 \pi)^4 \delta^{(4)} ( P' - P - p - q ) \sum |M|^2 \\
	 &=& G_F^2 \cdot \frac{2}{\pi^3} \cdot \int_{0}^{(\Delta - m_e)} 
				p ( \Delta - \frac{p^2}{2M} - x ) \;dx  \cdot ( x \sqrt{m_{\nu}^2 + x^2} )
\end{eqnarray}
	Here $p$ is the electron momentum, related to $x$ by
\begin{equation}
	p^2  \;=\; \frac{(\Delta - x)^2 - m_e^2}{1 + (\Delta - x)/M}
\end{equation}

	Because Chodos et al\cite{tachyon-chodos} did not attempt a canonical quantization of pseudo-Dirac 
	equation (\ref{chodos-equation}), their expression for the tritium decay spectrum is missing
	a contribution from the transient modes in a full tachyon field theory.  As we will show in this paper, 
	this contribution appears as an excess at the end-point of the tritium $\beta$-decay spectrum. (See Fig. 
	\ref{spectrum total}.)  
	
\begin{figure}[hbtp]
\begin{center}
\includegraphics[width=\textwidth]{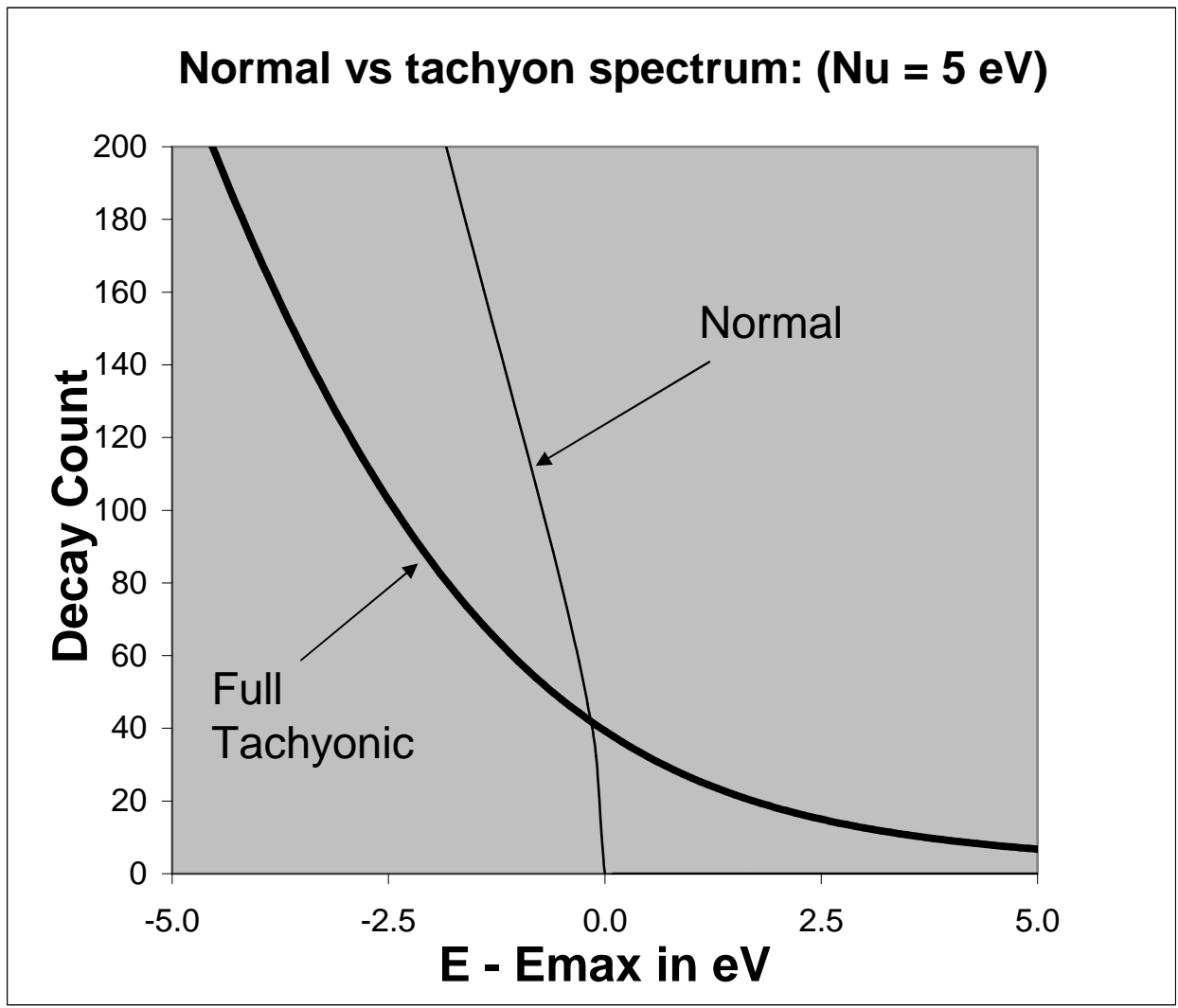}
\caption{Normal v full tachyonic Tritium $\beta$-decay.}
\label{norm v tach}
\end{center}
\end{figure}

	As a result of this transient contribution, the full tachyonic decay spectrum near the endpoint
	is no longer a straight line (see spacelike momenta contribution in Fig. \ref{spectrum total}), but a 
	concave curve, in contrast to a convex parabolic curve for a normal neutrino. (See Fig. \ref{norm v tach}.)

\section{Background}

	Earlier studies of tachyon field theory encountered many difficulties.
	It is well-known that a tachyon field $\psi_{_{sp}}$ containing only spacelike momenta 
	($\vec{q}\cdot \vec{q} > m^2$) does not obey micro-causality.  
      The equal time anti-commutator $\{ \psi_{_{sp}} (\vec{x}, 0), \psi^{\dagger}_{_{sp}}( \vec{y}, 0)\}$ 
	is not a simple spatial delta function, but has a space-like tail that destroys micro-causality.

	To make a spatial delta function, you need to include Fourier components with $\vec{q} \cdot \vec{q} < m^2$. 
	But these modes are complex solutions, and the problem becomes one of physical interpretation of exponential 
	decaying and runaway states.  Because there was no context for the inclusion or presence of these 
	``unphysical'' modes, the canonical quantization of a tachyon field theory has largely lain dormant over 
	the years.

	In a previous paper\cite{Chang}, we have reported on the complete canonical quantization of 
	tachyonic\cite{tachyon} (faster-than-light) neutrino field theory.  While considering Majorana 
	mass terms for the neutrino field, $\psi_{L}$, we found that by coupling it to a sterile negative-metric 
	$\psi^{\prime}_{L}$ field, the neutrino mass becomes space-like.  The Majorana mass-mixing term is a local 
	interaction, and therefore the resulting tachyonic field is a Fourier transform over {\it {\bf all}} momenta, 
	including the ``unphysical'' $\vec{q} \cdot \vec{q} < m^2$ momenta region.  We thus have a well-defined 
	context for the treatment of all Fourier components of the neutrino field.
	
	As shown in our previous paper\cite{Chang}, the $\vec{q} \cdot \vec{q} \;<\; m^2$
	are crucial to restoring the locality and microcausality of the resulting field theory. 
	They lead to a rich structure of the vacuum as a medium in which the neutrino 
	propagates with velocity exceeding that of light. 

	In this paper, we follow up on this work and study in particular the full implications of the
	tachyonic nature of the neutrino on the tritium $\beta$-decay spectrum, including these ``unstable'' 
	transient modes.

\section{Tachyonic Neutrino Field Theory}
	In ref\cite{Chang}, we considered the Majorana mass term for the $\psi_{L}$ neutrino field, 
	and coupling it to a sterile negative-metric $\psi^{\prime}_{L}$ field makes the neutrino mass 
	space-like.  
\begin{eqnarray}
	\gamma \;\cdot \;\partial \;\psi_{L} (\vec{x}, t)  &=& -\; m \;\gamma_2 \; 
				\psi^{\prime *}_{L} (\vec{x}, t) \label{Majorana-1} \\
	\gamma \;\cdot \;\partial \; \psi^{\prime}_{L} (\vec{x}, t) &=& + \; m \; \gamma_2 \; 
				\psi^{*}_{L} (\vec{x}, t)   \label{Majorana-2}
\end{eqnarray}
	The Lagrangian takes the form 
\begin{equation}
	L = - \;\bar{\psi_{L}} \gamma \cdot \partial \:\psi_{L} + \;\bar{\psi}^{\prime}_{L} \gamma \cdot \partial
		\:\psi {}^{\prime}_{L} \;- \; m \;\bar{\psi_{L}} \:C\: \bar{\psi_{L}^{\prime}} ^{T}
		\; - m \; \psi {}^{\prime} {}^{T}_{L} \:C\: \psi_{L}  \label{Lagrangian}
\end{equation}
	where $C$ is the charge conjugation matrix taken here to be $\gamma_2 \gamma_4$, and the $\gamma_{\mu}$
	matrices are all hermitian, with $\mu = 1, \ldots, 4$, and $\gamma_4 \equiv - i \gamma_o$.
	The negative-metric nature of the $\psi^{\prime}_{L}$ field manifests itself through the kinetic energy
	term of the Lagrangian.  The fields satisfy the canonical anti-commutation rules
\begin{eqnarray}
	\{  \psi_{L} (\vec{x}, t) \;,\;  \psi^{\dagger}_{L} ( \vec{y}, t) \}  &=&  \;\;\; \delta ( \vec{x} - \vec{y} ) \\
	\{  \psi {}^{\prime}_{L} (\vec{x}, t) \;,\;  \psi^{\prime \dagger}_{L} ( \vec{y}, t) \}  &=&   
			 -  \delta ( \vec{x} - \vec{y} ) 
\end{eqnarray}
	with all other anti-commutators vanishing.

	In the language of a composite four-component field,
	$\Psi$, where
\begin{equation}
	\Psi ( \vec{x}, t ) = \left( {\begin{array}{r}
					  \psi_{L} ( \vec{x}, t) \\
					-\; i \sigma_2 \psi^{\prime *}_{L} ( \vec{x}, t ) 
					      \end{array}} \right)  \label{Psi-field}
\end{equation}
	the Majorana equations of motion, eq. (\ref{Majorana-1},\ref{Majorana-2}), may be combined into a single 
	$4$-component equation which turns out to be the same as the pseudo-Dirac equation (eq. \ref{chodos-equation}) 
	first written down by Chodos, Hauser, and Kostelecky.\cite{tachyon-chodos}

	Note that the negative-metric $\psi^{\prime}_{L}$ does not participate in weak interactions.  Its only
	interaction is through the Majorana mass-mixing between $\psi_{L}$ and $\psi^{\prime}_{L}$ fields leading
	to the space-like mass of the neutrino.

\section{Canonical Quantization}
	A canonical quantization of this field theory reveals a rich structure of the vacuum.  
	The fully-dressed vacuum states are Nambu-Jona-Lasinio\cite{NJL} condensates of the $\psi_{L}$ field with 
	the negative-metric but sterile $\psi^{\prime}_{L}$.  

	Unlike the usual case, however, where the Hilbert space is built on a single physical vacuum 
	state, the Hilbert space here is built on two nilpotent vacua, which can be identified with the 
	scattering $in$ and $out$ states (see ref.\cite{Chang}),
\begin{equation} \left.{\begin{array}{lclcl}
	< \Phi_{in} \;\;| \Phi_{in} \;> &=& < \Phi_{out} \;| \Phi_{out} \;>  &=& 0 \\
	< \Phi_{out} \;| \Phi_{in} \;> &=& < \Phi_{in} \;\;| \Phi_{out} \;>   &=& 1 \\
			\end{array}} \right\}
\end{equation}
	The full Hamiltonian, $H$, is hermitian in the Hilbert space of $in$ and $out$ vacua.

	The propagators for the $\psi_{L}$ and $\psi^{\prime}_{L}$ fields may be summarized in terms of the 
	propagator for the $four$-component field, $\Psi$, of eq.(\ref{Psi-field})
\begin{equation}
	< \Phi_{out} | \; T \left( \Psi ( x ) \; \bar{\Psi} ( y ) \right) \; | \Phi_{in} >
			=  \int \frac{d^4 p}{(2 \pi)^4} \;\;\frac{- \gamma \cdot p + i \;m \gamma_{_5} }
				{p \cdot p  - m^2 - i\epsilon} \;\;{\rm e}^{i p \cdot ( x 
				- y) }  \label{propagator}
\end{equation}
	The Lorentz co-variance of this propagator is a direct consequence of the canonical anti-commutator
\begin{equation}
	< \Phi_{out} | \; \{ \Psi ( x ) \;,\; \Psi^{\dagger} ( y ) \; \} 
			\; | \Phi_{in} > |_{x_o = y_o} = \delta ( \vec{x} - \vec{y} ) 
\end{equation}
	so that micro-causality property is respected
\begin{equation}
	< \Phi_{out} | \; \{ \Psi ( x ) \;,\; \Psi^{\dagger} ( y ) \; \} 
			\; | \Phi_{in} >  = 0  \;\;\;{\rm for} \; (\vec{x} - \vec{y})^2 > 0
\end{equation}
	Here the metric is chosen such that
\begin{equation}
	\gamma_{\mu} \cdot p_{\mu}  \;\equiv\; \vec{\gamma} \cdot \vec{p} - \gamma_o \cdot p_o
\end{equation}

\section{Tritium $\beta$-decay lifetime}

\begin{figure}[hbtp]
\includegraphics[width=\textwidth,angle=-90]{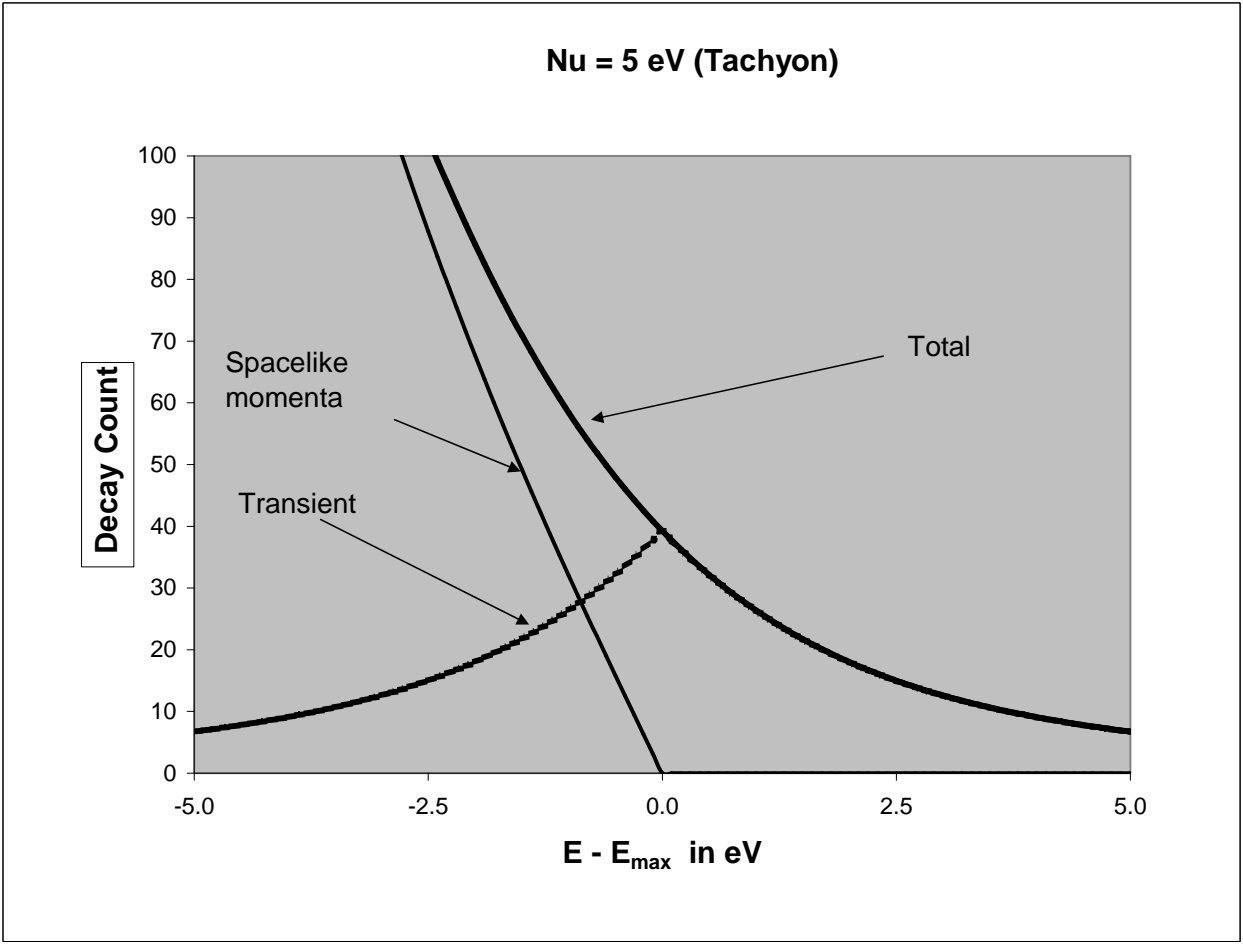}
\caption{Full tachyonic contribution to Decay.}
\label{spectrum total}
\end{figure}

	The effective interaction Hamiltonian responsible for tritium $\beta$-decay may be written
\begin{equation}
	H_{int}  \;=\;  \frac{G_F}{\sqrt{2}} \; \bar{\psi_{_{P}}} \gamma_{\mu} (1 + \gamma_{_5} ) \;\psi_{_N}
						\cdot	\bar{\psi_{e}} \gamma_{\mu} (1 + \gamma_{_5} ) \;\psi_{\nu}
						+ h.c.
\end{equation}
	By considering the perturbative expansion for the survival amplitude\footnote{\samepage We use $| \;;in, out >>$ to distinguish the full scattering $in$- and
	$out$- states from the $in$- and $out$- states in the pure neutrino sector.  The full Hamiltonian includes
	the weak interaction $H_{int}$.} 
	of the parent tritium nucleus 
	of momentum, $P'$,{}
\begin{equation}
	<< P' s' ;\;out |  P' s ;\;in >> \;=\; < P' s' ;\;out | \; T( e^{i \int H_{int} dt}) \; | P' s ;\;in >
\end{equation}
	we find as usual
\begin{equation}
	\tau  \;\delta_{s's} \;=\; \lim_{T \rightarrow \infty} \frac{1}{2T} \int d^4x d^4y  \;< P' s'; \;out |
				\; H_{int} (x) \; H_{int} (y) \;| P' s; \;in >
\end{equation}
	Here $M' -i/\tau$ gives the position of the pole in the tritium propagator.  
	
	Upon using the full neutrino propagator in eq.(\ref{propagator}), and performing all the space-time
	integrals, we find
\begin{eqnarray}
	\tau &=&  \frac{1}{2} \frac{1}{(2P'_o)} \int \;\frac{d^3P}{(2\pi)^3 (2P_o)}
					\frac{d^3p}{(2\pi)^3 (2p_o)}\frac{d^3q}{(2\pi)^3 (2q_o)} \;(2\pi)^3 \delta^{3}(P'-P-p-q) 
					\;\sum |M|^2 \nonumber \\
				&&	\left\{ (2 \pi)\delta(P'_o-P_o-p_o-q_o) \;\;\theta( \vec{q} \cdot \vec{q} - m_{\nu}^2 )\Bigl|_{q_o = \omega}
						\right.\nonumber \\
				&& \left.				+   \;\;\;\frac{2 \kappa}{(P'_o - P_o -p_o)^2 + \kappa^2}
						\;\;\;\;\theta(m_{\nu}^2 - \vec{q} \cdot \vec{q})\Bigl|_{q_o = - i \kappa}
									  \right\}  \label{full life-time}
\end{eqnarray}
	Here $\omega \equiv \sqrt{\vec{q}\cdot\vec{q} - m_{\nu}^2} \ge 0 $ is the physical neutrino energy for
	spacelike momenta while $\kappa \equiv \sqrt{m_{\nu}^2 - \vec{q}\cdot\vec{q}}$ is related to the inverse
	lifetime of the transient mode.  $\kappa$ ranges over the values
\begin{equation}
	m_{\nu} \;\ge\; \kappa  \ge 0
\end{equation}

	Eq.(\ref{full life-time}) gives the full contribution to the decay life-time due to the tachyonic pole in
	the propagator ($ q_{\mu} \cdot q_{\mu} - m_{\nu}^2 $).  The energy conserving delta function gives the
	familiar contribution due to physical spacelike neutrino momenta.  The new term arising from the transient
	modes is a Breit-Wigner like enhancement at $x=0$.  For very small $m_{\nu}$, this Breit-Wigner term 
	may be approximated by
\begin{equation}
	\frac{2 \kappa}{x^2 + \kappa^2}  \longrightarrow \; 2 \pi \delta (x)
\end{equation}
	However, for small but finite $m_{\nu}$, the transient contribution is not energy conserving, and so there
	is an excess even for $x < 0$, with a width of the order of $m_{\nu}$. (See fig. \ref{spectrum total} ).
	
	Upon performing the integration over neutrino momenta, the final expression for the lifetime takes the form
\begin{eqnarray}
	\tau  &=&  G_F^2 \cdot \frac{2}{\pi^3} \int_{-|x_{res}|}^{\Delta - m_e} dx \; p (\Delta - 
							\frac{p^2}{2M} - x)\left\{  x \cdot \sqrt{m_{\nu}^2 + x^2} \cdot \theta(x)\right. \nonumber \\
				& &	 \left. + \frac{m_{\nu}^2}{4} - \frac{m_{\nu}^4 |x|}{2(\sqrt{m_{\nu}^2 + x^2} - |x|)}
							\cdot \frac{1}{2 x^2 + m_{\nu}^2 + 2 |x| \sqrt{m_{\nu}^2 + x^2}} \right\}
\end{eqnarray}
	In this expression, the lower limit of the $dx$ integration depends on the experimental resolution 
	for $E_{max}$.  Electrons with $x$ below the lower limit,  $ - |x_{res}|$, would have energy
\begin{equation}
	E = \Delta - \frac{p^2}{2M} + | x_{res} | \equiv E_{max} + |x_{res}|
\end{equation}
	that is indistinguishable from background non-decay events, and are thus not included in the decay counting rate.
	
	To evaluate the total transient contribution to the tritium lifetime, we take the $m_{\nu} \rightarrow 0$
	approximation, and find
\begin{equation}
	\tau \bigl|_{transient} \;=\; \frac{2 G_F^2 }{3 \pi^2} \;p_{\max} \;(\Delta - \frac{p_{max}^2}{2M}) 
						\cdot m_{\nu}^3
\end{equation}
	where $p_{max} = \sqrt{\Delta^2 - m_e^2}/\sqrt{1 + \Delta/M}$.

\section{Conclusion}
	In this paper, we have included the transient modes in the tritium decay spectrum and shown how they lead to an excess of events near the endpoint.  The transient mode contribution is a Breit-Wigner like enhancement at the endpoint.  The effect of this excess is to `fill up' the linear slope associated with the physical spacelike momenta at the endpoint and make the resultant decay counting rate spill over into  `unphysical' electron energies, viz. those whose energies are greater than $E_{max}$.  From an experimental point of view, there is uncertainty in the precise mass of the tritium, and of the daughter helium nuclei, and so the spill over, within experimental resolution, is included in the decay counting rate.  

	Precision measurements of the tritium endpoint with these endpoint transient contributions in mind would settle not only the age-old question as to the mass of the neutrino, but also as to whether it is tachyonic or a normal particle.


\begin{thebibliography}{99}


\bibitem{Mainz}
	Ch. Weinheimer, B. Degen, A. Bleile, J. Bonn, L. Bornschein, O. Kazachenko, A. 
	Kovalik, E.W. Otten, \PLB {460}{219}{1999}, ``{\em High precision measurement of the tritium 
	$\beta$ spectrum near its endpoint and upper limit on the neutrino mass}''.  
	For latest update, see
	``{\em The neutrino mass direct measurements}'', Christian Weinheimer, hep-ex/0306057. 
	10th Int. Workshop on Neutrino Telescopes, Venice/Italy March 2003. 


\bibitem{Troitsk}
	V.M. Lobashev, V.N. Aseev, A.I. Belesev, A.I. Berlev, E.V. Geraskin, A.A. Golubev, 
	O.V. Kazachenko, Yu.E. Kuznetsov, R.P. Ostroumov, L.A. Rivkis, B.E. Stern, N.A. Titov, S.V. Zadorozhny, 
	Yu.I. Zakharov , ``{\em Direct Search for mass of neutrino and Anomaly in the Tritium Beta spectrum}'', 
	\PLB {460}{227}{1999}, \NPProc {91}{280}{2001}.


\bibitem{KATRIN}
	For update on the status of KATRIN, visit {\em http://www-ik.fzk.de/~katrin/index.html} 



















\bibitem{tachyon-chodos}
	A. Chodos, A. I. Hauser, V.A. Kostelecky, 
	\PLB {150}{431} {1985}.




\bibitem{Chang}
	N.P. Chang, \MPLA {16} {2129} {2001} ({\em hep-ph/0105153})



\bibitem{tachyon}
	G. Feinberg,
	\PRD {17} {1651} {1978}. See also
	M.P. Bilaniuk, V.K. Deshpande, E.C.G. Sudarshan, \AmJPhys {30}{718}{1962};
	J. Dhar, E.C.G. Sudarshan,
	\PhysRev {174} {1808}{1968}.








\bibitem{NJL}   
      Y. Nambu and G. Jona-Lasinio, \PhysRev {122}{345} {1961}; \ibid {124} {246} {1961}.









 









\end{thebibliography}
\end{document}